\documentstyle[12pt]{article}
\begin{document}
\title{DIMENSIONALITY AND FRACTALS}
\author{B.G. Sidharth$^*$\\
B.M. Birla Science Centre, Hyderabad 500 063 (India)}
\date{}
\maketitle
\footnotetext{$^*$E-mail:birlasc@hd1.vsnl.net.in}
\begin{abstract}
In this paper we first show that the usual three dimensionality of space,
which is taken for granted, results from the spinorial behaviour of
Fermions, which constitute the material content of the universe. It is
shown that the resulting three dimensionality rests on two factors which
have been hitherto ignored, viz., a Machian or holistic property and the
stochastic underpinning of the universe itself.\\
However the dimensionality is scale dependent in the sense that at very
large scales, or at very small scales, we encounter a different dimensionality, as
indeed is borne out by observation and experiment. For example the
large scale structures in the universe are cellular in nature on the one hand,
and we encounter fractional charges and handedness at very small scales.\\
Finally it is shown how fractal dimensions can emerge and as an illustrative
example it is shown how this could explain the magnetism of objects like
Planets on the one hand and White Dwarf stars and Pulsars on the other.
\end{abstract}
\section{Introduction}
In Classical Physics and in Quantum Mechanics including Quantum Field
Theory it is tacitly assumed that space time intervals can be made
arbitrarily small, even though Heisenberg's Uncertainity Principle then
dictates that we will have to deal with arbitrarily large momenta and
energies: Space time, whether it be Riemannian or Minkowski is treated as
a differentiable manifold. Furthermore, barring exceptional cases space time
is taken to have three plus one dimensionality.\\
However over the past 15 years an alternative viewpoint has been fruitfully
explored namely that the space time of micro physics is fractal
\cite{r1,r2,r3,r4}.
Specifically, it was pointed out that at the De Broglie wavelength and below,
there was a doubling of dimensionality. This again was very suggestive,
because this fractal dimension is characteristic of Brownian motion
\cite{r5} and the link with the various attempts at a stochastic
formulation of Quantum Mechanics was opened up \cite{r6,r7,r8,r9,r10}.\\
We will now describe a recent formulation of elementary particles that combines
stochastic and Quantum Theory to provide a pleasing rationale to the above
considerations on the one hand, and justifies and deduces from the theory
a picture consistent with data and observation.
\section{Quantum Mechanical Black Holes}
To start with let us first adhoc treat an electron as a Kerr-Newman Black
Hole. As is well known \cite{r11} this classical Kerr-Newman metric accurately
describes the field of an electron including the puzzling, purely Quantum
Mechanical gyro magnetic ratio $g = 2$. The price we have to pay for this
however is that, in classical terms there is a naked singularity: The horizon
of the Kerr-Newman Black Hole becomes complex \cite{r12},
\begin{equation}
r_+ = \frac{GM}{c^2} + \imath b, b \equiv (\frac{G^2Q^2}{c^8} +
a^2 - \frac{G^2M^2}{c^4})^{1/2}\label{e1}
\end{equation}
thus contradicting the censorship hypothesis which forbids naked singularities.
This is symptomatic of the fact that General Relativity and Quantum Mechanics
have been irreconcible.\\
On the other hand, the position coordinate for an electron as is well known
is given by \cite{r13}
\begin{equation}
x_\imath = (c^2p_\imath H^{-1} t + a_\imath ) + \frac{\imath}{2}
c\hbar (\alpha_\imath - cp_\imath H^{-1})H^{-1},\label{e2}
\end{equation}
where $a_\imath$ is an arbitrary constant with $c\alpha_\imath$
the velocity operator with
Eigen values $\pm c$. It is also well known that the real part in (\ref{e2}) is
the usual position while the imaginary part arises from the Zitterbewegung,
a rapid oscillation that is inexplicable in classical terms.\\
What is interesting is that in both (\ref{e1}) and (\ref{e2}) the imaginary part is
$\sim$ the Compton wavelength $\hbar/mc$. Equation (\ref{e1}) is inexplicable
in classical physics but arises naturally in Quantum Theory as in (\ref{e2}).
From a Quantum Mechanical point of view Dirac has explained the imaginary
or Zitterbewegung part by invoking the fact that according to the Heisenberg
Uncertainity Principle, measurement of space time points is unphysical.
Our physical measurements are averaged over intervals of atleast the
Compton scale. As we will see below the reconciliation of (\ref{e1}) and
(\ref{e2}) comes from a combination of fractal and stochastic approaches:
Indeed in Nelson's approach any particle is subjected
to an underlying Brownian motion described by forward and backward Wiener
processes, yielding the complex wave function of Quantum Mechanics, while
the doubling of the fractal dimension noted in Section 1 denotes a transition
from real to complex coordinates, as can be seen from (\ref{e2}) for example.\\
Before proceeding further it may be mentioned that (\ref{e1}) and (\ref{e2})
yield a reconciliation of General Relativity and Quantum Theory on the
one hand and a unification of Electromagnetism and Gravitation on the
other. This has been described in detail for example, in \cite{r14,r15},
the Kerr-Newman metric describing both the gravitational and
electromagnetic fields being symptomatic of this fact.\\
Our starting point is the well known Random Walk equation
\begin{equation}
R = l \sqrt{N}\label{e3}
\end{equation}
This is also the famous Eddington Large Number Relation, $R$ being the radius
of the universe, $N$ being the number of elementary particles, typically
pions and $l$ the pion Compton wavelength. Given (\ref{e3}) one can easily
deduce that $R=cT$, where $T$ is the age of the universe, and
\begin{equation}
T = \sqrt{N} \tau\label{e4}
\end{equation}
where $\tau = l/c$ is the pion Compton time.
In previous communications \cite{r16,r17,r18}
(\ref{e3}) was used
as the starting point for a stochastic formulation of Quantum Theory. We
will now indicate how the Dirac equation, and thereby a formulation for the
most elementary particles namely Quarks and Leptons comes from (\ref{e3}).\\
For this we first observe that while (\ref{e3}) gives us the Compton wavelength
of a typical elementary particle, (\ref{e2}) and even (\ref{e1}) yield the
Compton wavelength and time as a minimum cut off space time interval. Infact
it was shown that the Compton scale or equivalently quantized space time at
this scale could be considered to be more fundamental than Planck's energy
quanta \cite{r16} in the sense that the latter, which was a
starting point for Quantum Theory follows from the former. The concept of
discrete space time has been studied by several authors for over five
decades \cite{r19,r20,r21}. It can also
be shown to provide a rationale for a maximal velocity $c$, and so
Special Relativity itself \cite{r16,r22}. On the other hand
Snyder \cite{r19} showed that discrete space time is compatible with Special
Relativity and deduced
\begin{equation}
[x, y] = (\imath a^2 / \hbar)L_{z,}  [t, x] = (\imath a^2 / \hbar c)M_{x,}\label{e5}
\end{equation}
\begin{equation}
[x, p_x] = \imath \hbar [1+(a/\hbar)^2 p^2_x]\label{e6}
\end{equation}
and similar equations where $p^\mu$ denotes the four momentum and $a$ the
fundamental length and $L_x, M_x$ etc. have their usual meaning. Interestingly as $a \to 0$, we recover the
usual commutation relations of Quantum Theory. A coordinate shift in
Minkowski space then gives
\begin{equation}
\psi' (x_j) = [1 + \imath \epsilon (\imath \epsilon_{ljk} x_k \frac{\partial}
{\partial x_j}) + 0 (\epsilon^2)] \psi (x_j)\label{e7}
\end{equation}
Taking $a$ in (\ref{e5}) and (\ref{e6}) to be the Compton wavelength, one
can then show that
\begin{equation}
t =  \left(\begin{array}{l}
        1 \quad 0 \\ 0 \quad -1 \\
        \end{array}\right), \vec x =
   \ \  \left(\begin{array}{l}
         0 \quad \vec \sigma \\ \vec \sigma \quad 0 \\
       \end{array}\right)\label{e8}
\end{equation}
provides a representation for the coordinates $x$ and $t$. Substitution of
(\ref{e8}) in (\ref{e7}) then gives the Dirac equation \cite{r23}
\begin{equation}
(\gamma^\mu p_\mu - mc^2) \psi = 0\label{e9}
\end{equation}
Thus the Dirac equation is a consequence of quantized space time which itself
has its origins in the Random Walk equations (\ref{e3}) and (\ref{e4}).\\
The Quantum Mechanical spin half and the Compton scales symbolising
Zitterbewegung and the Brownian motion symbolised by (\ref{e3}) and (\ref{e4}),
as also the fractal doubling of dimensions referred to can all now be seen to
be closely interrelated.\\
We now observe that it is this spin half of Quantum Theory, that makes it
stand apart from classical physics. The big problem in the unification of
General Relativity and Quantum Mechanics is precisely this spinorial double
connectivity \cite{r11}. On the other hand it is also the spin half that leads in
a natural way to three dimensional space \cite{r11,r24}. But the above description and the three
dimensionality, it must be stressed, are valid for scales greater than the
Compton scale. We now recapitulate a description of quarks and neutrinos
in the above model. For this we observe that at the Compton scale itself
the negative energy two spinor
$\chi$ of the full four rowed Dirac spinor begins to
dominate. Moreover under reflections, $\chi$
behaves like a psuedo-spinor,
$$\chi \to - \chi$$
that is as a
density of weight $n = 1,$ so that \cite{r12},
\begin{equation}
\frac{\partial \chi}{\partial x^\mu} \to \frac{1}{\hbar} [\hbar \frac{\partial}
{\partial x^\mu} - n\hbar \Gamma_\sigma^{\mu \sigma}] \chi\label{e10}
\end{equation}
$\Gamma$'s being the usual Christoffel symbols.\\
We can easily identify the electromagnetic
four potential in (\ref{e10}).
The fact that $n = 1$ explains why the charge is discrete. We can also
immediately see the emergence of the metric tensor from the $\Gamma's$ (and the resulting
potential).\\
We now use the fact that the metric tensor $g_{\mu \nu}$ resulting from (\ref{e10})
satisfies an inhomogenous Poisson equation (Cf.\cite{r23} for details), whence
\begin{equation}
g_{\mu \nu} = G \int \frac{\rho u_\mu u_\nu}{|\vec r - \vec r'|}
d^3\vec r\label{e11}
\end{equation}
where now we require the volume of integration to be the Compton volume.
As shown elsewhere \cite{r12,r14,r15,r25},
given (\ref{e11}) which is also the linearized equation of General Relativity,
we can get a geometrized formulation of Fermions
leading to the Kerr-Newman metric and which explains the remarkable and supposedly
coincidental fact that the Kerr-Newman metric describes the field of an
electron including the anomalous gyro magnetic ratio $g = 2$.\\
All this was also
shown to lead to a unified description of electromagnetism, gravitation
and strong interactions \cite{r14,r25}.\\
We now show how a unified description of quarks and
leptons can be obtained from (\ref{e11}) and how the concept of fractal
dimensioinality is tied up with it.\\
From (\ref{e10}) and (\ref{e11}) we get \cite{r23}
\begin{equation}
A_0 = G \hbar \int \frac{\partial}{\partial t} \frac{(\rho u_\mu u_\nu )}
{|\vec r - \vec r'|} d^3r \approx \frac{ee'}{r}\label{e12}
\end{equation}
for $|\vec r - \vec r'| > >$ the Compton wavelength where $e' = e$ is the
test charge.\\
Further, from (\ref{e12}), as in the discrete space time case,
$d \rho u_\mu u_\nu = \Delta \rho c^2 = mc^2$ and $dt = \hbar/
mc^2$, we get
$$A_0 = \frac{e^2}{r} \sim G \frac{\hbar}{r} \frac{(mc^2)^2}{\hbar}$$
or
\begin{equation}
\frac{e^2}{Gm^2} \sim 10^{40} (\sim \sqrt{N})\label{e13}
\end{equation}
(\ref{e13}) is the well known but hitherto purely empirical relation expressing
the ratio of the gravitational and electromagnetic strengths here
deduced from theory.\\
If however in (\ref{e11}) we consider distances of the order of the Compton wavelength,
it was shown that we will get instead of (\ref{e12}), a QCD type potential
\begin{eqnarray}
4 \quad \eta^{\mu v} \int \frac{T_{\mu \nu} (t,\vec x')}{|\vec x - \vec x' |} d^3 x' +
(\mbox terms \quad independent \quad of \quad \vec x), \nonumber \\
+ 2 \quad \eta^{\mu v} \int \frac{d^2}{dt^2} T_{\mu \nu} (t,\vec x')\cdot |\vec x - \vec x' |
d^3 x' + 0 (| \vec x - \vec x' |^2) \propto - \frac{\propto}{r} + \beta r\label{e14}
\end{eqnarray}
where $T_{\mu \nu} \equiv \rho u_\mu u_\nu$. Equation (\ref{e14})
can lead to a reconciliation
of electromagnetism and strong interactions \cite{r25}. For this we need to
obtain a formulation for quarks from the above considerations. This
is what we will briefly recapitulate.\\
The doubleconnectivity or spin half of the electron as mentioned earlier
leads naturally to three dimensional space \cite{r11}. This however
breaks down at Compton scales and so we need to consider
two and one dimensions. Using the well known fact that
each of the $\rho u_\imath u_j$ in (\ref{e12}) is given by $\frac{1}{3} \epsilon$
\cite{r26}, $\epsilon$ being the energy density, it follows immediately
that the charge would be $\frac{2}{3} e$ or $\frac{1}{3}e$
in two or one dimensions, exactly as for quarks. At the same time as we are
now at the Compton scale, these fractionally charged particles
are confined as is expressed by the confining part of the QCD
potential (\ref{e14}). Further, at the Compton scale, as noted earlier
we encounter predominantly the negative energy components of the Dirac spinor
with, opposite parity. So these quarks would show neutrino type handedness,
which indeed is true.\\
Thus at one stroke, all the peculiar empirical characteristics of the quarks
for which as Salam had noted \cite{r27}, there was no theoretical rationale,
can now be deduced from theory. We can even get the correct order of
magnitude estimate for the quark masses \cite{r25}.
On the other hand neutrinos have vanishingly small mass. So their Compton wavelength is very large and by
the same argument as above, we encounter predominantly the negative energy
components of the Dirac spinor which have opposite parity, that is the
neutrinos display handedness.\\
Thus handedness and fractional charge are intimately tied up with dimensionality.\\
It is interesting that the above conclusion about quarks in a low dimensional
context can be deduce from the more conventional analysis of Laughlin \cite{r28}
of the fractional Quantum Hall Effect \cite{r29}. (In this context, it must
be mentioned that fractional charges have recently been experimentally
detected \cite{r30}. Laughlin's magnetic length now becomes the Compton
wavelength (Cf.\cite{r28}) and quarks show up as Laughlin's fractionally charged
quantum fluid of quasi electrons. Infact if in Laughlin's expression for total
energy (Cf.\cite{r29})
$$- \frac{1}{2} \pi^{1/2} e^2/r$$
we replace $r$ by the Compton wavelength and the charge $e$ by $e/3$ as seen
above, and equate this energy with $m_ec^2,$ , the electron energy and as
$r \sim$ of the Compton wavelength, we can deduce that $m \sim 10^3m_e$ that is
the constitutents of the quantum fluid show up with quark masses, their
fractional charge and handedness.
\section{Fractal Behaviour at De Broglie Scales and Low Dimensions}
As noted by Nottale and others \cite{r1}, fractal behaviour should appear at
De Broglie scales. We will now demonstrate this fact in the context of an
assembly of Fermions below the Fermi temperature in which case we are
at de Broglie scales, and in low dimensions,
from the point of view of the more conventional  non relativistic
theory.\\
We first observe that as is known \cite{r31} an assembly of Fermions below the
Fermi temperature occupies each and every single particle level, and this
explains the fact that it behaves like a distribution of Bosonic phonons
\cite{r32}: The Fermions do not enjoy their normal degrees of freedom.\\
It is known that \cite{r31} the energy density $\epsilon$ below the Fermi
energy is given by
\begin{equation}
\epsilon \propto \int^{p_F}_o \frac{p^2}{2m} d^3p\label{e15}
\end{equation}
where $p_F$ is the Fermi momentum. We can verify that the integral in
(\ref{e15}) is $\propto T_F^{2.5}$, where $T_F$ is the Fermi temperature,
so that we have,
\begin{equation}
\epsilon \propto T_F^{2.5}\label{e16}
\end{equation}
If we compare (\ref{e16}) with the result for $n$ dimensions \cite{r33,r34}
viz.,
$$\epsilon \propto T^{n+1}_F$$
we conclude that the assembly behaves as if it has a fractal dimensionality
$1.5$. This explains the Bosonization or semionic effect.\\
Interestingly, this could also provide a rationale for a Cooper type pair of
superconductity: Given the above fractal dimensionality, and using the fact
as seen in Section 2, that the charge per dimensions is $e/3$, it follows
that the charge of the Fermions would be $e/2$. So a pair of such charges
would be required for the usual charge $e$.\\
Another way to see this is that in the case of the Coulumb potential for
example the energy levels would be given by
$$E_n = - |E_n | = - \frac{\mu Z^2e^4}{8\hbar^2n^2},$$
rather than
$$E_n = - \frac{\mu Z^2e^4}{2\hbar^2n^2},$$
as in conventional theory owing to the fact that the $e^2$ of the electron is
replaced by $e^2/4$. Effectively $n$ is replaced by $2n$.\\
We now come to low dimensional behaviour. In one dimension
the assembly behaves as if
it were below the Fermi temperature, whatever be the temperature\cite{r32}. Indeed this
low temperature Quantum nature has recently been experimentally verified with
carbon nanotubes \cite{r35}.\\
We can come to a similar conclusion in the two dimensional case also. In this
case, for the Fermi temperature we can easily adapt the usual three dimensional
theory (cf.ref.\cite{r31}) to two dimensions to obtain
\begin{equation}
\frac{1}{\lambda^2} = \frac{1}{v}\label{e18}
\end{equation}
where $v \equiv \frac{V}{N}.$ Now using the expression for total energy in
two dimensions, viz.,
$$NkT \approx U = \frac{V}{2\pi \hbar^2} \int \frac{p^3}{2m} \langle n_p \rangle dp,$$
and invoking a partial integration and using the well known fact that the
derivative of $\langle n_p \rangle$ is a delta function peaked at $p_F$ \cite{r31}
we get
\begin{equation}
\frac{1}{\lambda} \approx \left(\frac{1}{v}\right)^{1/2}\label{ex}
\end{equation}
Equation (\ref{ex}) is the same as equation (\ref{e18}). However in this case we have worked
at an arbitrary temperature $T$. (It may be mentioned that we encounter the Quantum
behaviour and two dimensionality at the interface between two semiconductors).\\
To sum up, there would be bosonization or semionic behaviour below the Fermi
temperature or in low dimensions - that is at the fractal De Broglie scale.\\
Let us now consider an assembly of $N$ electrons. As is known, if $N_+$ is the
average number of particles with spin up, the magnetisation per unit volume
is given by
\begin{equation}
M = \frac{\mu (2 N_+ - N)}{V}\label{e19}
\end{equation}
where $\mu$ is the electron magnetic moment. At low temperatures, in the usual
theory, $N_+ \approx \frac{N}{2}$, so that the magnetisation given in
(\ref{e19}) is very small. On the other hand, for Bose-Einstein statistics we
would have, $N_+ \approx N$. With the above semionic statistics we have,
\begin{equation}
N_+ = \beta N, \frac{1}{2} < \beta < 1,\label{e20}
\end{equation}
If $N$ is very large, this makes an enormous difference in (\ref{e19}).
Let us use (\ref{e19}) and (\ref{e20}) for the case of Neutron stars.\\
In this case, as is well known, we have an assembly of degenerate electrons
at temperatures $\sim 10^7K$, whereas the Fermi
temperature is $\sim 10^{11}K$ (cf.for example \cite{r31}). So the above considerations
apply. In the case of a Neutron star we know that the number
density of the degenerate electrons, $n \sim 10^{31}$ per c.c. \cite{r36} So using (\ref{e19}) and (\ref{e20}) and remembering
that $\mu \approx 10^{-20}G$ (Gauss), the magnetic field near the Pulsar is $\sim
10^{11}G \leq 10^8$ Tesla, as required.\\
Some White Dwarfs also have magnetic fields. If the White
Dwarf has an interior of the dimensions of a Neutron star, with a similar
magnetic field, then remembering that the radius of a White Dwarf is about
$10^3$ times that of a Neutron star, its magnetic field would be $10^{-6}$
times that of the neutron star, which is known to be the case.\\
It is quite remarkable that the above mechanism can also explain the magnetism
of the earth \cite{r37}. As is known the earth has a solid core of radius of about
1200 kilometers and temperature about 6000 K \cite{r38}. This core is made up
almost entirely of Iron $(90 \%)$ and Nickel $(10 \%)$. It can easily be
calculated that the number of particles $N \sim 10^{48}$, and that the
Fermi temperature $\sim 10^5 K$. In this case we can easily verify using
(\ref{e19}) and (\ref{e20}) that the magnetic field near the earth's surface
$\sim 1G$, which is indeed the case.\\
It may be mentioned that the anomalous Bosonic behaviour given in (\ref{e20}) would imply a
sensitivity to external magnetic influences which could lead to effects
like magnetic flips or reversals.\\
To see this, we observe that the number of electrons, with spin aligned along
a magnetic field $B$ which is introduced, where,
$$B < < \epsilon_F/2\mu,$$
is given by (cf.ref. \cite{r31}), using Fermi-Dirac statistics,
$$N_+ \approx \frac{N}{2} (1+\frac{3\mu B}{2\epsilon_F})$$
That is, $\beta$ in (\ref{e20}) is given by
$$\beta \approx \frac{1}{2},$$
and the introduction of the field $B$, does not lead to a significant magnetic
field in (\ref{e19}). But, if as we have seen, $\beta \ne \frac{1}{2}$, but
rather, $\beta > \frac{1}{2}$, then in view of the fact that $N$ is very
large, the contribution from (\ref{e19}) could be significant.\\
Indeed in the case of the earth,
magnetic reversals do take place from time to time and are as yet not
satisfactorily explained \cite{r39}.\\
Remembering that the core density of Jupiter is of the same order as that
of the earth, while the core volume is about $10^4$ times that of the earth,
we have in this case, $N \sim 10^{52}$, so that the magnetization $MV$, from
(\ref{e20}) is $\sim 10^4$ times the earth's magnetism, as required\cite{r38}.\\
\section{Discussion}
In the light of the preceding considerations we now make the following
observations:\\
1. As mentioned earlier all of contemporary physics, from General Relativity
to Quantum Field Theory assumes a space time continuum, but as pointed out
by Nottale\cite{r1}, scale relativity has been ignored, even though
Heisenberg's Uncertainity Principle attributes arbitrarily large momenta
and energy for arbitrarily small space time intervals. Infact it has been
shown that \cite{r40} if there is a break in scale, then we
recover the Compton wavelength. As pointed out the fractal dimensions of
space time at the Compton scale have been deduced by Nottale and Ord.
In the nonrelativistic case the fractal dimensionality begins at the
De Broglie wavelength as suggested by the work of Abbott-Wise(Cf.ref.\cite{r1}).
However, at larger scales, there is scale independence. The minimum
space time intervals referred to earlier express precisely this fact.
At very large scales again, the structures of the universe display cellular
and filamentary characteristics symptomatic of fractality\cite{r41}.\\
If these minimum space time cut offs are ignored, as pointed out elsewhere
\cite{r12,r18}, we encounter the divergences of Quantum Field
Theory. On the other hand, once they are recognized the naked singularities
of General Relativity get fudged and disappear as seen in Section 2 and
what Wheeler calls the greatest crisis of Physics is overcome, and also
the irreversibility of time, via the Kaon experiment for example
is explained\cite{r18}.\\
As T.D. Lee and others have observed \cite{r20,r21} the space time continuum
is but an approximation. Unphysical effects like divergences, Zitterbewegung,
reversed time or negative energies or chaotic spin flips which appear in a
generalization of the stochastic formulation to relativistic Quantum Mechanics
are all symptomatic of this approximation being ignored.\\
It is also interesting to note that the minimum space time cut off,
be it at the De Broglie scale in the nonrelativistic case or the Compton scale
in the relativistic case involves the mass. Further it has been noted by
Nottale and others, that  Quantum Mechanical properties like spin and the wave
function can be attributed to this fractal structure \cite{r1}. It has
also been shown that properties like mass and spin result from the space
time cut offs\cite{r12}.\\
2. In previous communications\cite{r12,r16} a pleasing correspondence
was seen between the Quantum Mechanical Black Holes of Section 2 on the one hand, and
the De Broglie-Bohm hydromechanical formulation, as also Nelson's stochastic
formulation on the other hand. The former follows from the latter if we
consider the scale of the Compton length and time in which case there is a
doubling of coordinates which is precisely expressed by the fractal dimension
being double the usual dimension, as referred to earlier. This is also the
reason for the puzzling and inexplicable fact that Newman's imaginary shift
of coordinates leads from the classical Kerr metric to the Kerr-Newman metric
which is Quantum Mechanical as seen in Section 2. This complexification is
again symptomatic of the fractal doubling of dimensions. However as observed
by Nottale \cite{r1} all this should not be construed as the hidden
variable theory, because of the stochastic and unpredictable underpinning.\\
3. We would like to re-emphasize that the fractality of space time in micro
physics, at Compton scales implies that the usual concept of an iron clad
space time has to be replaced by the concept of a Brownian space time heap
the Compton scale being, as seen above, the cut off mean path length or
size. To put it in a pictorial manner, this is like the Richardson scenario
\cite{r5} of delineating a jagged coastline with a thickened smooth curve,
the thickness being the Compton scale akin to the mean free path of Wheeler's
travelling salesman as shown elsewhere \cite{r23}. However it should be pointed out that this picture is Machian or
holistic, as indeed is expressed by equations like (\ref{e3}) and (\ref{e4}),
or other large number relations of cosmology that can be deduced from the
theory \cite{r42} like Weinberg's so called mysterious and empirical relation that
expresses the pion mass in terms of the Hubble Constant.

\end{document}